\documentclass[prl,aps,twocolumn,twoside,english,superscriptaddress]{revtex4}

\usepackage{epsfig,verbatim}
\usepackage{graphicx}
\usepackage{amsmath}
\usepackage{revsymb}

\newcommand{\ket}[1]{|  #1 \rangle}
\newcommand{\bra}[1]{ \langle #1  |}
\newcommand{\proj}[1]{\ket{#1}\!\bra{#1}}

\newcommand{\Tr}{{\mathrm{Tr}}}

\newcommand\bea{\begin{eqnarray}}
\newcommand\eea{\end{eqnarray}}
\newcommand{\beq}{\begin{equation}}
\newcommand{\eeq}{\end{equation}}

\begin{document}

\title{\Large Can closed timelike curves or nonlinear quantum mechanics improve quantum state discrimination or help solve hard problems?}
\author{Charles H. Bennett}\email{bennetc@watson.ibm.com}
\affiliation{IBM T.J. Watson Research Center, Yorktown Heights, NY
10598, USA}
\author{Debbie Leung}\email{wcleung@iqc.ca}
\affiliation{Institute for Quantum Computing, University of Waterloo, Waterloo, Ontario, Canada. N2L 3G1.
}
\author{Graeme Smith}\email{gsbsmith@gmail.com}
\affiliation{IBM T.J. Watson Research Center, Yorktown Heights, NY
10598, USA}
\author{John A. Smolin}\email{smolin@watson.ibm.com}
\affiliation{IBM T.J. Watson Research Center, Yorktown Heights, NY
10598, USA}

\date{\today}

\begin{abstract}
We study the power of closed timelike curves (CTCs) and other
nonlinear extensions of quantum mechanics for
distinguishing nonorthogonal states and speeding up
hard computations. If a CTC-assisted computer is presented with a
labeled mixture of states to be distinguished---the most natural
formulation---we show that the CTC is of no use.  The apparent contradiction with recent claims that CTC-assisted computers can perfectly
distinguish nonorthogonal states is resolved by noting that 
CTC-assisted evolution is nonlinear, so the output of such a computer on a
mixture of inputs is not a convex combination of its output on the
mixture's pure components. Similarly, it is not clear
that CTC assistance or nonlinear evolution help solve
hard problems if computation is defined as we recommend, as
correctly evaluating a function on a labeled mixture of orthogonal
inputs.
\end{abstract}

\maketitle

{\em Introduction:} Physicists and science fiction writers have long been
interested in time travel, wherein a person or object travels
backward in time to interact with a younger version of itself.
The many studies of such closed timelike curves have led to the general conclusion that,
while conditions for their creation may not arise in typical astrophysical or cosmological settings,
in principle there seems to be no barrier to their existence\cite{MTY88,Gott91,DJH92,Hawking92,Ori05}.

In the context of quantum computation, the most widely accepted model
of time travel, due to Deutsch~\cite{Deutsch91}, involves a
unitary interaction $U$ of a causality-respecting (CR) register with
a register that traverses a CTC. The physical states of
Deutsch's theory are the density matrices of quantum mechanics,
but the dynamics are augmented from the usual linear evolution.  For each
initial mixed state $\rho_{CR}$ of the CR register, the CTC register
is postulated to find a fixed point $\rho_{CTC}$ such that
\begin{equation}\Tr_{CR}(U\rho_{CR}\otimes\rho_{CTC}U^\dagger)=\rho_{CTC}.\label{Eq:DeutschFP}
\end{equation}
The final
state of the CR register is then defined in terms of the fixed point
as
\begin{equation}
\rho'_{CR}=\Tr_{CTC}(U\rho_{CR}\otimes\rho_{CTC}U^\dagger).
\label{Eq:DeutschEvol}
\end{equation} The induced mapping $\rho_{CR}\rightarrow\rho'_{CR}$
is nonlinear because the fixed
point $\rho_{CTC}$ depends on the initial state $\rho_{CR}$. The
nonlinear evolution leads to various puzzling consequences considered below, but, because the fixed point is
allowed to be a mixed state, it always exists \cite{Deutsch91}, thereby avoiding the
notorious ``grandfather paradox'' wherein some initial
conditions lead to no consistent future \cite{foot}.

\begin{figure}

(a) \includegraphics[height=1.2in]{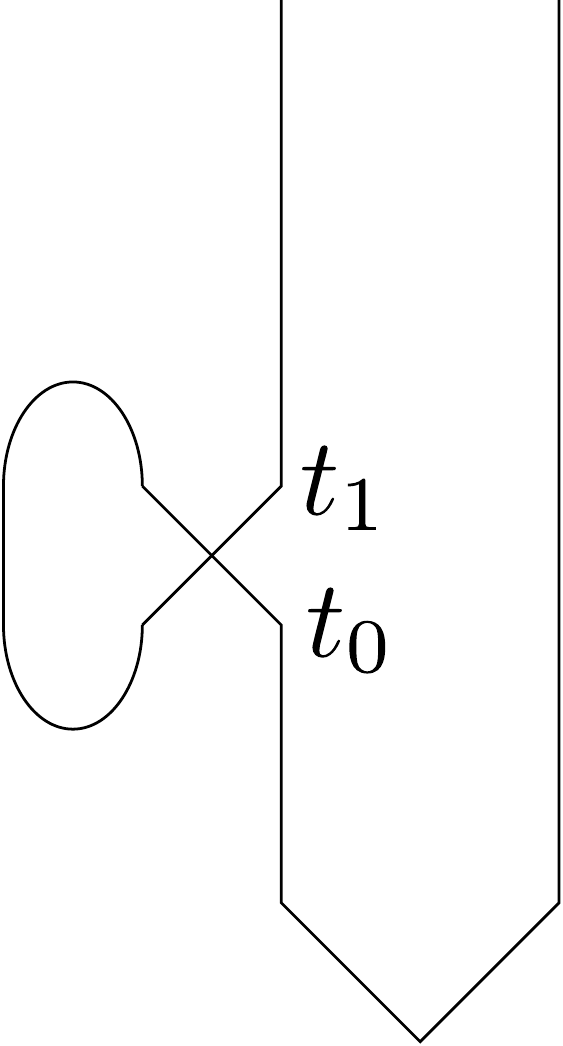} \hspace{.1in} (b) \includegraphics[height=1.2in]{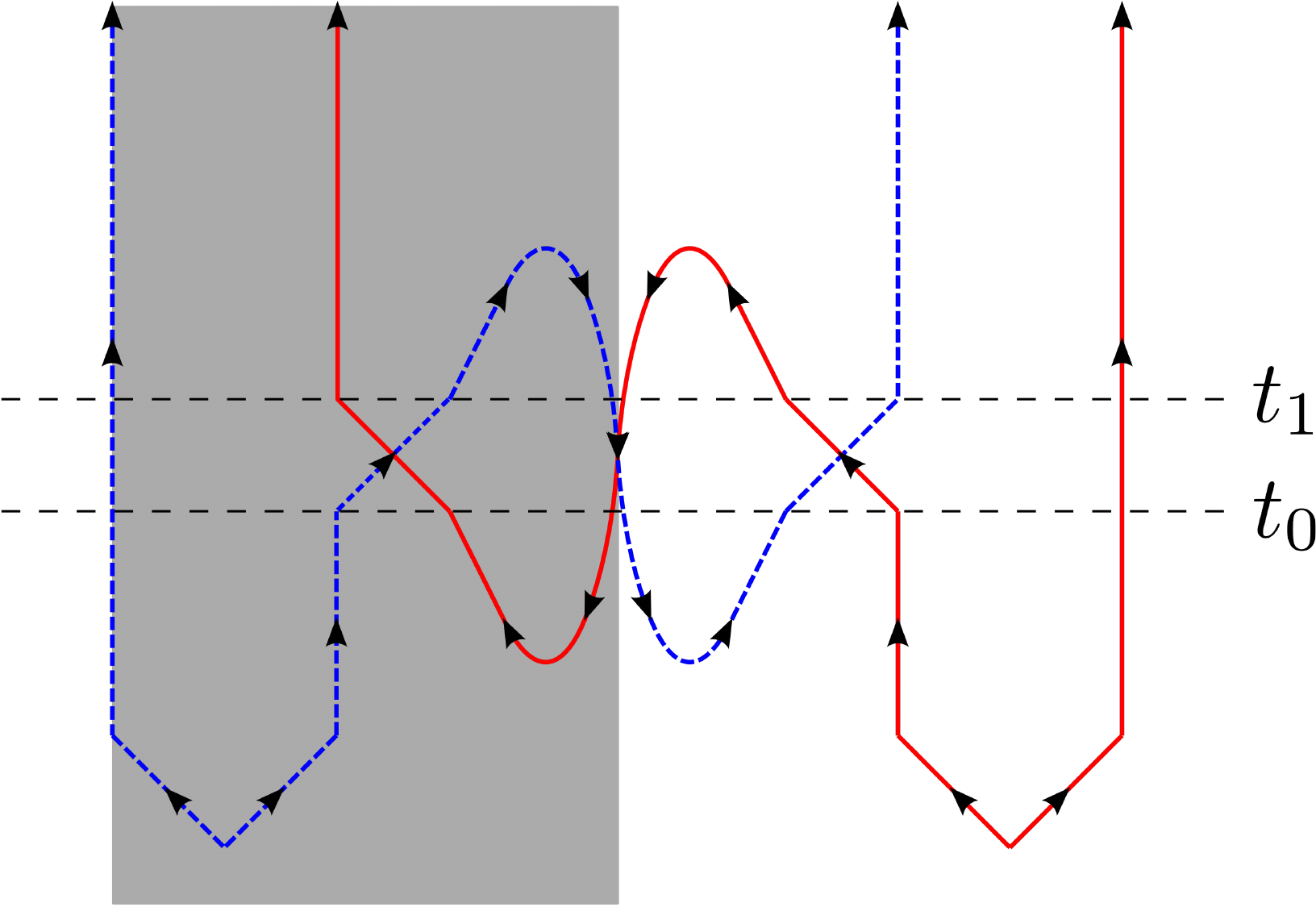}

\caption{Sending half of an EPR pair along a CTC. (a) Single universe
  picture. An EPR pair is created in the distant past.  At time $t_0$
  a qubit emerges from the CTC and at time $t_1$ half of the EPR pair
  is put into the CTC. According to Deutsch's prescription the density
  matrix of the CTC system at $t_0$ is equal to the CTC density matrix
  at $t_1$.  Nevertheless the joint state at any time after $t_1$ is a
  product state.  (b) Multiple universe picture.  In both universes an
  EPR pair is created in the distant past.  At time $t_0$ a qubit
  emerges from the CTC in each universe.  At time $t_1$ in each
  universe half of an EPR pair is put into the CTC and goes back in
  time to emerge at $t_0$ in the other universe.
  Each EPR particle originally created is entangled with a partner in
  the other universe and in a product state with the other particle in
  its own universe.} \label{Fig:EPR}
\end{figure}

In Deutsch's model, the mixed-state fixed point $\rho_{CTC}$
explicitly begins in a product state with the $CR$ register.
Thus, the universe may evolve from a pure to mixed
state, which is not normally allowed by quantum mechanics.  To
recover a pure state picture Deutsch appeals to the
multiverse of the many-worlds interpretation, where the CTC system
in our world is entangled with other worlds'
CTC and CR systems. This kind of
mixed state runs counter to the ``church of the larger Hilbert space''
philosophy applicable to CTC-free quantum mechanics,
which views mixed states as always being subsystems of larger
entangled pure systems in {\em this} universe.

To illustrate Deutsch's model, consider putting half
of a maximally entangled state into a CTC (FIG. \ref{Fig:EPR}).
There are now two causality respecting qubits, $A$ and $B$,
and a single CTC qubit.  The unitary of Eq.~(\ref{Eq:DeutschFP})
is the swap operation between CTC and B.  Finding the fixed
point gives $\rho_{CTC} = \frac{1}{2}I$, which along with
Eq.~(\ref{Eq:DeutschEvol}) gives a final state of $\rho^\prime_{AB} =
\frac{1}{4}I_A\otimes I_B$ on the causality respecting qubits.
Strangely, not only does the CTC cause an evolution from a pure
to mixed state but the simple act of sending $B$ along a CTC disentangles
it from $A$.  A pure state is recovered
by considering both our initial universe and the universe
with which the CTC interacts.

{\em Distinguishing States:} Our work is motivated by \cite{BHW09},
which explored the benefits of CTCs for
state discrimination.  There it was shown that for
any pair of pure states, $\ket{\phi_0}$ and $\ket{\phi_1}$, there is
a CTC-assisted circuit that maps these to orthogonal states
$\ket{0}$ and $\ket{1}$, respectively.  This was interpreted as
distinguishing nonorthogonal states, an
impossibility in standard quantum mechanics. It was also shown that
the linearly dependent set $\{\ket{0},\ket{1}, \ket{+},\ket{-}\}$,
where $ \ket{\pm}= \frac{1}{\sqrt{2}}(\ket{0}\pm\ket{1})$, can be
mapped to four orthogonal states using a CTC.  Interpreting this as
distinguishing these states leads us to a truly remarkable
conclusion: a CTC can be used to distinguish $I/2 =
\frac{1}{2}\proj{0} + \frac{1}{2}\proj{1}$ from $I/2 =
\frac{1}{2}\proj{+} + \frac{1}{2}\proj{-}$.  Apparently a
CTC lets us distinguish {\em identical} states.  Of course, it
is not entirely clear what this means!

The authors of \cite{BHW09} knew something had gone awry, and
speculated that either their own analysis or Deutsch's model
must be wrong.  To resolve this conundrum, we look more closely
at what it means to discriminate among quantum states.
Discrimination is necessarily {\em adversarial}, in the sense that a
referee, Rob, presents the discriminator, Alice, with a system
prepared in an unknown state $\ket{\phi_0}$ or $\ket{\phi_1}$.  Before
Rob gives her this system she does not know which state he will
prepare, but after some processing she should be able to tell Rob
whether it was $\ket{\phi_0}$ or $\ket{\phi_1}$. Since Rob will choose
the state according to some physical process and must remember his
choice in order to check that Alice has succeeded, the joint state of
Alice and Rob before any distinguishing operation is
\begin{equation}
\rho_{RA} = \sum_{x=0}^1 p_x \proj{x}_R \otimes \proj{\phi_x}_A.
\label{Eq:Rsystem}
\end{equation}
Alice will now apply some operation to the $A$ system. We will say
she has succeeded if the joint state afterwards is
\begin{equation}\rho'_{RA}=\sum_{x=0}^1 p_x \proj{x}_R \otimes \proj{x}_{A}.
\label{Eq:rhoprime}
\end{equation}

Our formulation of the problem may seem obvious, and even a bit
pedantic, but as we will now see it has major consequences for the
power of CTCs: they are entirely useless for state
discrimination.  To see this, suppose we have a CTC-assisted
protocol for distinguishing $\ket{\phi_0}$ and $\ket{\phi_1}$ that
takes a causality-respecting input $A$ and closed timelike
curve register CTC.  The causality respecting region consists
of $R$ and $A$, with the fact that Alice does not have access to $R$
reflected in the restriction of Eq.~(\ref{Eq:DeutschFP}) to $U = I_R \otimes V_{A,CTC}$.
Even without access to a CTC, because she knows
$p_x$ and $\ket{\phi_x}$ (though not the particular value of $x$)
she can solve the fixed point problem (\ref{Eq:DeutschFP})
to get $\rho_{CTC}$.
So, she can prepare a quantum state $\rho_{CTC}$ and, given a state to
distinguish on $A$, apply $V$ to the joint $ACTC$ system and generate
the same output state $\rho_{RA}^\prime$ as if she actually had a CTC.
In short, Alice can simulate the help of a CTC by solving the
fixed-point problem herself, eliminating any advantage the CTC may
have offered.

\begin{figure}
\includegraphics[width=2.6in]{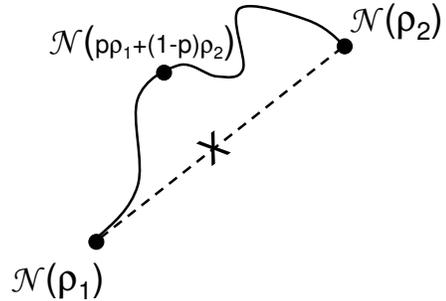}
\caption{The linearity trap.  The action of a nonlinear map $\cal N$
  on states $\rho_1$ and $\rho_2$ does not determine the
  action on their mixture. An example of such a map is the evolution
  of states in the CTC model.  So, although a CTC allows
  nonorthogonal pure states to be mapped to orthogonal outputs this
  does not suffice to identify the states in an unknown mixture.
  Similarly, the apparent power of CTC assisted computations is not
  enough to allow a user to sample the correct output of the
  computation over an arbitrary distribution of inputs.}
\label{fig:trap}
\end{figure}

\begin{figure}
\includegraphics[width=2.6in]{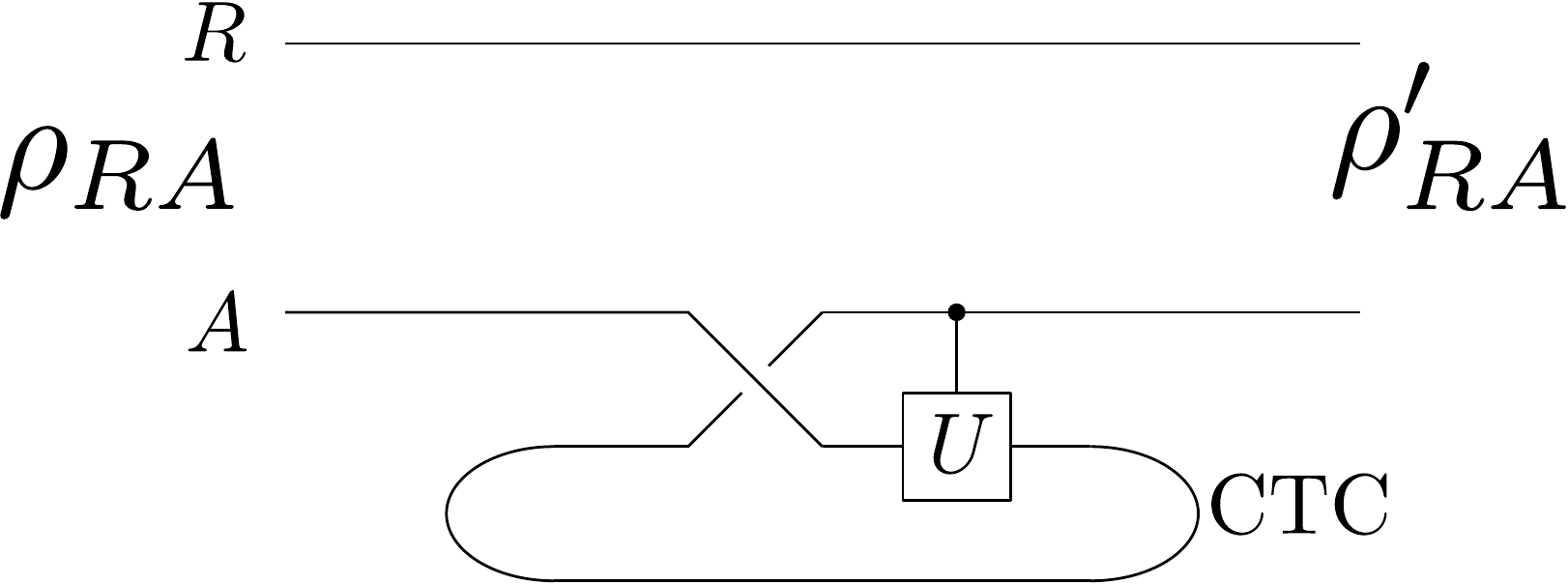}
\caption{The state discrimination circuit of \protect\cite{BHW09}.
  The circuit on $A$ and CTC is designed to distinguish pure states
  $\ket{0}$ and $\ket{\psi}$.  $U$ is chosen with
  $U\ket{\psi}=\ket{1}$ which leads to fixed points $\proj{0}$
  when $\ket{0}$ is input and $\proj{1}$ when $\ket{\psi}$ is input.
  However, faced with the task of distinguishing an unknown
  mixture labeled by $R$ as in Eq. (\protect\ref{Eq:Rsystem}) the
  output $\rho'_{RA}=\rho_R \otimes \rho'_A$.  The output of the
  circuit is independent of the identity of the state.}
\label{fig:CTC2states}
\end{figure}

How do we reconcile the fact that CTCs do not improve state
discrimination with the finding of \cite{BHW09} that any pair of
pure states can be mapped to orthogonal outputs using a CTC?  We
must be careful to avoid falling into the following ``linearity
trap'': while in standard quantum mechanics the evolution of a
mixture is equal to the corresponding mixture of the evolutions of
the individual states, in a nonlinear theory this is not generally true (see
Fig. \ref{fig:trap}). Thus, while the circuit of \cite{BHW09} (see
Fig. \ref{fig:CTC2states}) can map $\ket{x}_R\ket{\phi_x}_A
\rightarrow \ket{x}_R\ket{x}_A$ it does not map the mixed state Eq.
(\ref{Eq:Rsystem}) to the desired output (\ref{Eq:rhoprime}) but
rather to
\begin{equation}
\left(\sum_{x=0}^1 p_x \proj{x}_R\right)\otimes \rho_A'
\label{eq:5}.
\end{equation}
The output $\rho_A'$ depends on the ensemble $\{p_x,\ket{\phi_x}\}$
 but not on the particular value of $x$.  Indeed, even when presented with
a superposition of states, $\sum_x \sqrt{p_x}\ket{x}_R\ket{\phi_x}_A$, the circuit fails.  The correlations between
$R$ and $A$ are completely broken, reflecting the disentangling
nature of Deutsch's model of CTCs.

{\em Computational consequences:} We now focus on the
computational power of closed timelike curves.  Several authors have
concluded that access to CTCs would have substantial
computational benefits.  For example, \cite{Brun02} suggested that
a CTC would allow a classical computer to efficiently factor
composite numbers and gave hints that a
CTC-enhanced computer may be much stronger.  In \cite{Bacon04} it was argued
that a CTC-assisted quantum computer could efficiently
solve NP-complete problems, a feat widely believed impossible
for a quantum computer alone.  The strongest results about
computation using CTCs are those of \cite{AW08}, where it
was reported that the power of a polynomial time bounded computer
(either classical or quantum) assisted by CTCs is exactly PSPACE, the
class of problems that can be solved in a space polynomial in the
problem size but potentially exponential time.  Because PSPACE is
thought to contain many problems that cannot be solved efficiently
without CTC assistance, this would suggest that CTCs are extremely
useful for computation.

Analyzing the power of CTCs is a subtle business, as we saw with
state discrimination.  To understand what is going on it's
useful to spell out exactly what we mean by ``computing.''  We first
have to ask how the input to the calculation is chosen.  If it is
chosen by some physical process, the inputs have some probability
distribution that depends on the selection procedure.  As we have
seen, for a nonlinear theory the performance of a circuit depends
on the probability distribution over the inputs.  So, probably the
strongest form of computation would be to provide the correct answer
for every input distribution.

In \cite{AW08} (and implicitly in \cite{Bacon04}) the class ${\rm
  BQP}_{\rm CTC}$ is defined, where BQP stands for {\bf B}ounded error
{\bf Q}uantum {\bf P}olynomial time and the subscript refers to its
augmentation by a CTC.  By their definition, an algorithm
succeeds if it gives the correct answer on every pure state input.
In fact, in all previous work on computation with CTCs it is shown
that for a fixed pure state input (and even for {\em all} pure state
inputs) the proposed circuit reaches the correct output.
However, to argue that it follows that a physical computer would work on
every input distribution would be to fall prey to the linearity trap.

It is easy to check that the circuits of \cite{Bacon04,AW08}
for computing a function $F(x)$, when applied to a uniform mixture of
inputs (with an external referee remembering which one has been
supplied), do not generate the state
\begin{equation*}
\frac{1}{X}\sum_{x=1}^X \proj{x}_R \otimes \proj{F(x)}_A
\end{equation*}
but give a product state similar to Eq.~(\ref{eq:5}).
The output of the circuit is uncorrelated
with the input.  Thus, we believe claims
that a quantum computer with CTC assistance can efficiently
solve NP-complete and PSPACE-complete problems are dubious, at least
for the natural definition of computation as the ability to
find the correct output no matter how the input is chosen.

Given their definition of ${\rm BQP}_{\rm CTC}$, the arguments in
\cite{Bacon04,AW08} are valid, but this definition
is problematic because it implicitly limits the computer to operating
on a single input rather than a range of possible inputs. The
physical interpretation of a single input might be that one has made
a firm and unwavering decision to use a CTC to solve a particular
problem (e.g. whether black has a winning strategy in Go), rather
than a class of problems, as is usual in computational complexity
theory. This decision may as well be taken to have existed since the
beginning of time, and cannot depend on any other part of the universe.
Only then will the CTC-assisted computer give
the desired result. There is no physical problem with this, as it is
equivalent to the universe having been created with special objects
containing answers to particular questions, but it is not very
appealing in terms of the common meaning of computation. For
example, one might be disappointed by a Go computer claiming to
know the winner of the standard 19x19 game but unable to shed any
light on variants using boards of other sizes.

Thus we suggest a new complexity class ${\rm BQPP}_{\rm CTC}$, whose
definition is identical to that of ${\rm BQP}_{\rm CTC}$ of
\cite{AW08}, except that the computer must produce correctly correlated
mixtures of input-output pairs for all labeled input distributions (and the input is supplied as a string rather
than a circuit).  We do not know whether ${\rm BQPP}_{\rm CTC}$ is stronger than the
unassisted BQP.  Since the CTC fixed point is uncorrelated with the
inputs to a circuit, it seems like a fairly weak resource, akin to
``quantum advice'' \cite{Advice1,Advice2}. Fortunately, the argument
in \cite{AW08} that ${\rm BQP}_{\rm CTC} \subset {\rm PSPACE}$ holds
for our definition of computing, so at least we know that ${\rm
BQPP}_{\rm CTC}$ is in PSPACE.

Similar arguments apply to classical complexity classes like
P and BPP in the presence of CTCs.  If computation is
defined in the natural manner we recommend, CTCs have not been shown to
enlarge any of these classes.

{\em General Nonlinear Theories:}
Weinberg has proposed a general
approach for adding nonlinearities to quantum mechanics \cite{Wein89,WeinPRL89}.  It
was argued almost immediately that the theory has pathological properties.
Notably, \cite{Pol91,Gisin89} suggested
that the theory gives
faster than light communication. Moreover, modification to eliminate
this problem gives communication between branches of
the wavefunction, dubbed the ``Everett Phone''\cite{Pol91}.  It was also argued \cite{Peres89}
that it violates the
second law of thermodynamics.  Finally, \cite{AL98} argued
that {\em any} nonlinear version of quantum mechanics allows
the efficient solution of NP-complete and $\#$P-complete problems.

In the follow-up work to \cite{Wein89,WeinPRL89}---the instantaneous
communication of \cite{Pol91,Gisin89}, the second law violation
of \cite{Peres89} and the computational speed-up of
\cite{AL98}---the arguments proceed by considering
the evolution of some pure state, then inferring the induced evolution
of their mixture.  It is the linearity trap again!  For example, just
as the CTC-circuits of \cite{Bacon04,AW08} fail on a mixed
state, the circuits of \cite{AL98}  using the nonlinearities of \cite{Wein89,WeinPRL89,Pol91} give outputs that
are uncorrelated with their inputs when applied to a labeled mixture,
resulting in no computational speed-up.  Because the linearity trap is so enticing,
we propose a rule of thumb for dealing with nonlinear theories:

\noindent{\bf The Principle of Universal Inclusion:}
{\em The evolution of a nonlinearly evolving system may
depend on parts of the universe with which it does not interact.}

This principle reflects the fact that 1) calculations ignoring any part
of the universe invite the linearity trap, and 2) theories
formulated only on subsystems are incomplete.  The parts of
the universe that are perilous to ignore in the nonlinear theories
above are the systems used to select inputs to
computational or information theoretic problems.  In linear quantum mechanics this causes
no problems because for an evolution ${\cal N}$ we have
\begin{equation}
I \otimes {\cal N} (\sum_i p_i \proj{i} \otimes \phi_i) =
\sum_i p_i \proj{i} \otimes {\cal N} (\phi_i)
\end{equation}
but in other theories this is not so.  Perhaps this is what Polchinski\cite{Pol91} was driving at in his discussions of the ``Everett
phone,'' cautioning against ``treat[ing] macroscopic systems as though
they begin in definite macroscopic states'' instead of considering their
entire histories.

{\em Discussion:} Much of the apparent power of CTCs and nonlinear
quantum mechanics comes from analyzing the evolution of pure states,
and extending these results linearly to find the evolution of mixed
states.  However, because mixed states do not have unique
decompositions into pure states this does not give an unambiguous
rule for evolution.  Indeed, the very nature and meaning of mixed
states may be ill defined in such theories. One could potentially
resolve this problem by including additional degrees of freedom
identifying the ``correct'' decomposition of mixed states, which
would restore the power of CTCs. Unfortunately, this resolution does
not reduce to standard quantum mechanics far from any CTC. We find it more
rewarding
to concentrate on theories that do, such as
Deutsch's formalism.  In such theories we can, far from the CTCs,
unambiguously define initial and final mixed states for the
the tasks of state discrimination and computation. We then find
that CTCs do not seem to help much in their accomplishment.

Besides \cite{Deutsch91,Wein89,WeinPRL89,Pol91}, there are several
models for CTCs and nonlinear quantum mechanics
\cite{Svet09,Hartle94,Politzer94,Hawking95,CD02,RMD09}. Their
information processing power is not known, and our work underscores
the necessity of clear and well-motivated definitions of the tasks
under consideration in any such study.

The reported pathological behavior of nonlinear quantum
mechanics could have been construed as explaining why nature
chose standard linear quantum mechanics.  Our
findings that many of these behaviors do not survive careful scrutiny
suggest that a well behaved nonlinear theory may be
possible.  In fact, as pointed out in \cite{Pol91} we could in
principle have large nonlinearities in a global theory that have
little or no consequence for experiments on small systems.  It would
have been nice to rule out nonlinearity by
causality or a prohibition on computational extravagance, but it seems
that we cannot.

{\em Acknowledgments:}
DL thanks CRC, CFI, ORF, NSERC, CIFAR,
and QuantumWorks.  CHB, GS, and JAS acknowledge DARPA QUEST contract HR0011-09-C-0047.  We are grateful to D. Bacon, T. Brun, C. Fuchs, R. Spekkens, and J. Watrous
for very helpful discussions.

\bibliographystyle{apsrev}

\begin{thebibliography}{26}
\bibitem[{foo()}]{foot}
\bibinfo{note}{Svetlichny \cite{Svet09} (cf. also \cite{GP03}) discusses a
  teleportation-based alternative to Deutsch's picture, in which the earlier
  turning point of CTC register is replaced by Bell state initial condition and
  the later turning point by a projective measurement onto the same Bell state.
  We plan to compare the two approaches in a future paper.}
\end{thebibliography}

\end{document}